\documentstyle[aps,epsf]{revtex}
\twocolumn

\newcommand{\infig}[2]{\begin{center}\mbox{\epsfxsize #2
\epsfbox{#1}}\end{center}}

\begin{document}
\input {epsf}

\newcommand{\beq}{\begin{equation}}
\newcommand{\eeq}{\end{equation}}
\newcommand{\beqa}{\begin{eqnarray}}
\newcommand{\eeqa}{\end{eqnarray}}

\def\ov{\overline}
\def\onlyif{\rightarrow}

\def\openone{\leavevmode\hbox{\small1\kern-3.8pt\normalsize1}}

\def\a{\alpha}
\def\b{\beta}
\def\g{\gamma}
\def\r{\rho}
\def\minus{\,-\,}
\def\eks{\bf x}
\def\kay{\bf k}

\def\ket#1{|\,#1\,\rangle}
\def\bra#1{\langle\, #1\,|}
\def\braket#1#2{\langle\, #1\,|\,#2\,\rangle}
\def\proj#1#2{\ket{#1}\bra{#2}}
\def\expect#1{\langle\, #1\, \rangle}
\def\trialexpect#1{\expect#1_{\rm trial}}
\def\ensemblexpect#1{\expect#1_{\rm ensemble}}
\def\kpsi{\ket{\psi}}
\def\kphi{\ket{\phi}}
\def\bpsi{\bra{\psi}}
\def\bphi{\bra{\phi}}

\def\ditto{\rule[0.5ex]{2cm}{.4pt}\enspace}
\def\th{\thinspace}
\def\ni{\noindent}
\def\thirty{\hbox to \hsize{\hfill\rule[5pt]{2.5cm}{0.5pt}\hfill}}

\def\set#1{\{ #1\}}
\def\setbuilder#1#2{\{ #1:\; #2\}}
\def\Prob#1{{\rm Prob}(#1)}
\def\pair#1#2{\langle #1,#2\rangle}
\def\Id{\bf 1}

\def\dee#1#2{\frac{\partial #1}{\partial #2}}
\def\deetwo#1#2{\frac{\partial\,^2 #1}{\partial #2^2}}
\def\deethree#1#2{\frac{\partial\,^3 #1}{\partial #2^3}}

\newcommand{\xx}{{\scriptstyle -}\hspace{-.5pt}x}
\newcommand{\yy}{{\scriptstyle -}\hspace{-.5pt}y}
\newcommand{\zz}{{\scriptstyle -}\hspace{-.5pt}z}
\newcommand{\kk}{{\scriptstyle -}\hspace{-.5pt}k}
\newcommand{\sx}{{\scriptscriptstyle -}\hspace{-.5pt}x}
\newcommand{\sy}{{\scriptscriptstyle -}\hspace{-.5pt}y}
\newcommand{\sz}{{\scriptscriptstyle -}\hspace{-.5pt}z}
\newcommand{\sk}{{\scriptscriptstyle -}\hspace{-.5pt}k}

\def\openone{\leavevmode\hbox{\small1\kern-3.8pt\normalsize1}}

\title{Quantum Cryptography using entangled photons in 
energy-time Bell states}
\author{
W. Tittel, J. Brendel, H. Zbinden, and N. Gisin
\\
\small
{\it Group of Applied Physics, University of Geneva, CH-1211, Geneva 4,
Switzerland}}
\maketitle

\abstract{We present a setup for quantum cryptography based on photon 
pairs in
energy-time Bell states and show its feasability in a laboratory experiment. 
Our scheme combines the advantages of 
using photon pairs instead of faint laser pulses and the 
possibility to preserve energy-time
entanglement over long distances. 
Moreover, using 4-dimensional energy-time states,
no fast random change of bases is required in our setup : Nature itself 
decides whether to measure in the energy or in the time base.}

\vspace{0.5 cm}
\normalsize

Quantum communication is probably one of the most rapidly growing 
and most exciting fields 
of physics within the last years \cite{physworld}. Its most mature
application is quantum cryptography (also called quantum key 
distribution), 
ensuring the distribution of a secret key between two parties. This key 
can be used afterwards  
to encrypt and decrypt secret messages using the one 
time pad \cite{Welsh}. 
In opposition to the mostly used "public key" systems 
\cite{Welsh}, the security of 
quantum cryptography is not based on
mathematical complexity but on an inherent property of 
single quanta. 
Roughly speaking, since it 
is not possible to measure an unknown quantum system without modifying it,
an eavesdropper manifests herself by introducing errors 
in the transmitted data. During the last years, several prototypes 
based on faint 
laser pulses mimicking single photons, have been 
developed, demonstrating that 
quantum cryptography not only works inside the laboratory, but in the 
"real world" as well \cite{physworld,plug&play,expquantumcryptography}. 
Besides, it has been shown that two-photon entanglement
can be preserved over large distances \cite{Longdistbell}, 
especially when being entangled in energy and time \cite{fullengthbell}. 
As pointed out by Ekert in 1991 \cite{Ekert91}, 
the nonlocal correlations engendered by such states can also be 
used to establish 
sequences of correlated bits at distant places, the advantage compared 
to systems based on
faint laser pulses being the smaller vulnerability against 
a certain kind of
eavesdropper attack \cite{QNDattack,security}.

Besides improvements in the domain of quantum key distribution, 
recent experimental progress in generating,  
manipulating and measuring the so-called Bell-states \cite{Bellstates}, 
has lead to fascinating applications like 
quantum teleportation \cite{teleportation}, dense-coding \cite{densecoding}
and entanglement swapping 
\cite{entanglementswapping}. 
In a recent paper, we proposed and tested a novel source for quantum 
communication 
generating a new kind of Bell states based on energy-time 
entanglement \cite{newsource}. 
In this paper, we present a first 
application, exploiting this new source for quantum cryptography. 
Our scheme 
follows Ekert's 
initial idea concerning the use of photon-pair correlations. 
However, in opposition, it implements Bell states and 
can thus be seen in the 
broader context of quantum communication. 
Moreover, the fact that energy-time entanglement 
can be preserved over long distances renders our source particulary 
interesting for long-distance applications.


To understand the principle of our idea, we look at Fig. 1.
A short light pulse emitted at time $t_0$ enters an interferometer having a 
path length difference which is large compared to the duration of the pulse.
The pulse is thus split into two pulses of smaller amplitudes, 
following each other with a fixed phase relation. The light is then 
focussed into a nonlinear crystal where some of the pump photons are  
downconverted into photon pairs.
Working with pump energies low enough to ensure that generation of 
two photon pairs
by the same as well as by two subsequent pulses
can be neglected, a 
created photon pair is described by 
\begin{eqnarray}
\ket{\psi}=\frac{1}{\sqrt2}\bigg(\ket{s}_A\ket{s}_B + 
e^{i\phi} \ket{l}_A\ket{l}_B\bigg).
\label{Bellstate}
\end{eqnarray}
$\ket{s}$ and $\ket{l}$ denote a photon created by a pump photon having 
traveled via the short or 
the long arm of the interferometer, and the indices A, B label the photons. 
The state (\ref{Bellstate}) is composed of only two discrete 
emission times and not
of a continuous spectrum.
This contrasts with the energy-time entangled states used up to now
\cite{Ekert91,Franson89}.
Please note that, depending on the phase $\phi$, 
Eq. (\ref{Bellstate}) describes two 
of the four Bell states. Interchanging $\ket{s}$ and $\ket{l}$ 
for one of the 
two photons leads to generation of the remaining two Bell-states. 
In general, the coefficients describing the amplitudes of 
the $\ket{s}\ket{s}$ and $\ket{l}\ket{l}$ states
can be different, leading to nonmaximally entangled states. However, in this
article, we will deal only with maximally entangled states.
Behind the crystal, the photons are separated and are sent to Alice and Bob, 
respectively (see Fig.1).
There, each photon travels via another interferometer, 
introducing exactly the same difference of travel times through one or 
the other arm
as did the previous
interferometer, acting on the pump pulse. If Alice looks at the arrival 
times of the photons 
with respect to the emission time of the pump pulse $t_0$ --
note that she has two detectors to look at--, she 
will find the photons in one of three time slots.
For instance, detection of a photon in
the first slot corresponds to "pump photon having traveled via the short 
arm and 
downconverted photon via the short arm". To keep it short, we refer to 
this process as 
$\ket{s}_P;\ket{s}_A$, 
where $P$ stands for the pump- and $A$ for Alice's photon. 
However, the characterization of the complete photon pair
is still ambiguous, since, 
at this point, the path of the photon having 
traveled to Bob (short or long in his interferometer) 
is unknown to Alice. 
Fig. 1 illustrates all processes leading to a detection in the different 
time slots both at 
Alice's and at Bob's detector. Obviously, this reasoning holds
for any combination of two detectors.
In order to build up the secret key, 
Alice and Bob now publicly agree about the events where both 
detected a photon 
in one of the satellite 
peaks -- without revealing in which one -- or both in the central 
peak -- without 
revealing the detector. This additional information enables both of 
them to know
exactly via which arm the sister photon, detected by the other person, 
has traveled. 
For instance, to come back to the above given example, if
Bob tells Alice that he detected his photon in a satellite peak as well, 
she knows that
the process must have been 
$\ket{s}_p; \ket{s}_A \ket{s}_B$. The same holds for Bob who now knows 
that Alice photon 
traveled via the short arm in her interferometer.
If both find the photons in the right peak, the process was
$\ket{l}_p; \ket{l}_A \ket{l}_B$. 
In either case, Alice and Bob have correlated detection times. 
The cross terms
where one of them detect a photon in the left and the other one
in the right satellite peak do not occur.
Assigning now bitvalues 0 (1) to the short (long) processes, Alice and Bob 
finally end up with a sequence of correlated bits. 

Otherwise, if both find the photon in
the central slot, the process must have been
$\ket{s}_p; \ket{l}_A \ket{l}_B$ or $\ket{l}_p; \ket{s}_A \ket{s}_B$. If
both possibilities are indistinguishable, we face the 
situation of interference and
the probability for detection by a given combination of 
detectors (e.g. the
"+"-labeled detector at Alice's and the "--" labeled one at Bob's) 
depends on 
the phases $\a$, $\b$ and $\phi$ in the three interferometers. 
The quantum mechanical treatment leads 
to $P_{i,j}~=~\frac{1}{2}\big( 1+ijcos(\a+\b-\phi)\big)$
with $i$,$j$ = $\pm1$ denoting the detector labels \cite{newsource}. 
Hence, chosing appropriate phase settings, 
Alice and Bob will always find perfect correlations in the output ports. 
Either both detect the photons 
in detector "--" (bitvalue "0"), or both in detector "+" (bitvalue "1"). 
Since the correlations depend on the phases and thus on the energy of the 
pump, signal and idler photons, we refer to this base as the energy base 
(showing wave like behaviour), 
stressing the complementarity with the other, the time basis 
(showing particle like behaviour). 
\begin{figure}
\infig{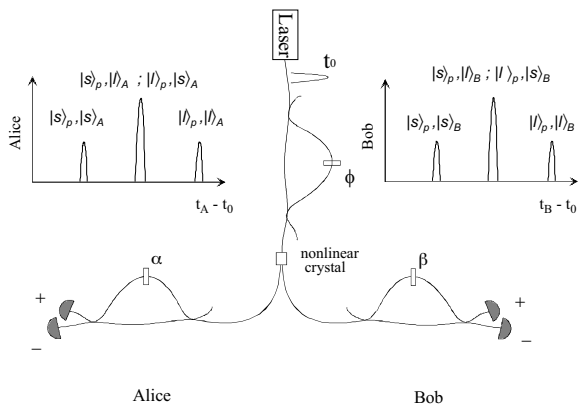}{0.9\columnwidth}
\caption{Schematics of quantum key distribution using 
energy-time Bell states.}
\end{figure}

Like in the BB84 protocol \cite{BB84}
it is the use of complementary bases that ensures the detection 
of an eavesdropper \cite{eavesBB84}. 
If we consider for instance the most intuitive intercept/resend 
strategy, the eavesdropper intercepts the photons, measures 
them in one of the two bases and sends new, accordingly 
prepared photons instead. 
Since she never knows in which basis Bob's measurement will take place, 
she will in half of the cases eavesdrop and resend the 
photons in the "wrong basis"
and therefore will statistically introduce errors in Bobs results, 
revealing in turn her presence. For a more general treatment of quantum 
key distribution and eavesdropping using energy-time complementarity,
we refer the reader to \cite{highalphabet}.


To generate the short pump pulses, we use a pulsed diode laser 
(PicoQuant PDL 800), 
emitting 600ps (FWHM) pulses  of 
655 nm wavelength at 
a repetition frequency of 80 MHz. The average power is of $\approx$ 10 mW, 
equivalent to an energy of 125 pJ per pulse.  
The light passes a dispersive prism, preventing the small quantity 
of also emitted infrared light to enter the subsequent setup, and 
a polarizing beamsplitter (PBS), serving as optical isolator.
The pump is then focussed into a singlemode fiber and is 
guided into a fiber optical
Michelson interferometer made of a 3 dB fiber coupler and chemically 
deposited silver end mirors. The path length difference corresponds to a 
difference of travel times of $\approx$ 1.2 ns, splitting the pump 
pulse into two,
well seperated pulses. The arm-length difference of the whole 
interferometer 
can be controlled using a pizoelectric actuator in order to ensure
any desired phase difference. Besides, the temperature is maintained stable.
In order to control the evolution of the polarization in the fibers, 
we implement three
fiber-optical polarization controllers, each one consisting of three 
inclinable fiber 
loops -- equivalent to three waveplates in the case of bulk optic.  
The first device is placed before the interferometer and ensures
that all light, leaving the Michelson interferometer by the input port
will be reflected by the already mentioned PBS and thus will 
not impinge onto the laser diode. The second controller serves to 
equalize the evolution of 
the polarization states within
the different arms of the interferometer, and the last one enables 
to control the 
polarization state  
of the light that leaves the interferometer by the second output port. 
The horizontally polarized light is now focussed into a 4x3x12 mm 
$KNbO_3$ crystal,
cut and oriented to ensure degenerate collinear phasematching, 
hence producing photon pairs
at 1310 nm wavelength -- within the so-called second 
telecommunication window. 
Due to injection losses of the pump into the fiber and losses 
within the interferometer, the average power 
before the crystal drops to $\approx$ 1 mW, and the energy per 
pulse -- remember that each initial pump pulse is now split into two --
to $\approx$ 6 pJ. The probability for creation of more than one photon 
pair within the 
same or within two subsequent pulses is smaller than 1 \%, 
ensuring the assumption that lead to 
Eq. \ref{Bellstate}. Behind the crystal, the red pump light
is absorbed by a filter (RG 1000). The downconverted photons 
are then focussed into 
a fiber coupler, separating them in half of the cases, and are guided
to Alice and Bob, respectively. The interferometers (type Michelson) 
located there have been 
described in detail in  \cite{fullengthbell}. They consist of a 3-port 
optical circulator, 
providing access to the second output arm of the interferometer, a 
3 dB fiber coupler and 
Faraday mirrors in order to compensate any birefringence within 
the fiber arms. 
To controll their phases, the temperature  
can be varied or can be maintained stable. Overall 
losses are about 6 dB. The
path length differences of both interferometers are equal with respect 
to the coherence length of the downconverted photons -- 
approximately 20 $\mu$m.
In addition, the travel time difference is the same than the one
introduced by the interferometer acting on the pump pulse. In this case, 
"the same" refers to
the coherence time of the pump photons, around 800 fs 
or 0.23 mm, respectively.

To detect the photons, the output ports
are connected to single-photon counters -- passively quenched 
germanium avalanche photodiodes, operated in
Geiger-mode and cooled to 77 K \cite{fullengthbell}. We operate 
them at dark count rates 
of 30 kHz, leading to quantum efficiencies of $\approx$ 5 \%. 
The single photon detection 
rates are of 4-7 kHz, the discrepancy being due to different 
detection efficiencies and losses in the circulators.
The signals from the detectors as well as signals being coincident with 
the emission of a pump pulse 
are fed into fast AND-gates. 


To demonstrate our scheme, we first measure the correlated events 
in the time base. 
Conditioning the detection at Alice's and Bob's detectors
both on the left satellite peaks,
($\ket{s_p}$,$\ket{s_A}$ and $\ket{s_p}$,$\ket{s_B}$, respectively) 
we count the 
number of coincident detections between both AND-gates, that is the
number of triple coincidences between emission of the pump-pulse 
and detections at Alice's and Bob's.
In subsequent runs, we measure these rates
for the 
right-right ($\ket{l_p}$,$\ket{l_A}$ AND $\ket{l_p}$,$\ket{l_B}$) 
events, as well as for the 
right-left cross terms. 
We find 
values around 1700 coincidences per 100 sec for the correlated, 
and around 80 
coincidences for the
non-correlated events (table 1). 
From the four times four                 
rates --remember that we have four pairs of detectors--, 
we calculate the different quantum bit error rates QBER, which is the
ratio of wrong to detected events. 
We find values inbetween 3.7 and 5.7 \%,
leading to a mean value of QBER for the time base of (4.6 $\pm$0.1)\%.

In order to evaluate the QBER in the energy base, we condition the 
detection at Alice's 
and Bob's on the central peaks. Changing the phases in any of the 
three interferometers, we observe interference fringes in the triple 
coincidence count 
rates (Fig.2). 
Fits yield visibilities of 89.3 to 94.5 \% for the different detector 
pairs (table 2). In the case of appropriatly chosen phases,
the number of correlated events are  
around 800 in 50 sec, and the number of errors are around 35. From these 
values, we calculate 
the QBERs for the four detector pairs. We find values inbetween 
2.8 and 5.4 \%, leading to a mean QBER for the energy base of 
(3.9$\pm$0.4) \%.
Note that this experiment can be 
seen as a Franson-type test of Bell inequalities 
as well \cite{Franson89,Bell}. From the mean visibility of 
(92.2$\pm$0.8)\%,
we can infer to a violation of Bell inequalities by 27 
standard deviations.

\begin{figure}
\infig{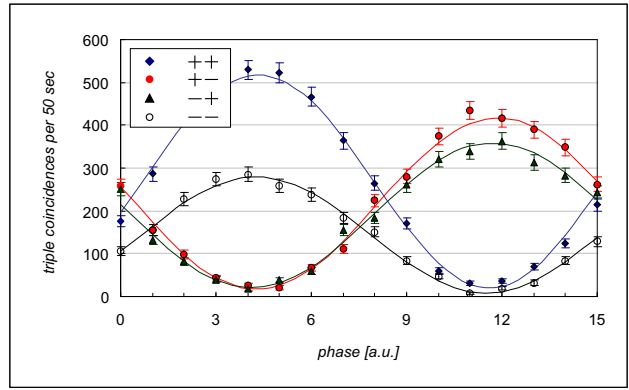}{0.95\columnwidth}
\caption{Results of the measurement in the energy basis. 
The different mean values are due to different detector efficiencies}
\label{fig2} 
\end{figure}

Like in all experimental quantum key distribution schemes, 
the found QBERs are non-zero, even in the absence of any eavesdropping. 
Nevertheless, they are still small enough to guarantee the detection of an 
eavesdropper attack, allowing thus secure key distribution. 
inequality 
The remaining $\approx$ 4\%
are due to accidental coincidences from uncorrelated events at the 
single photon detectors, 
a not perfectly localized pump pulse, non-perfect time resolution of 
the photon detectors, 
and, in the case of the energy basis, 
non-perfect interference. Note that the last mentioned errors
decrease at the same rate as the number of correlated events when 
increasing the 
distance between Alice and Bob (that is when increasing the losses). 
In contrast to that, the number
of errors due to uncorrelated events (point 1) stays almost constant 
since it is
dominated by the detector noise. Thus, 
the QBER increases with distance, however, only at a small rate since 
this contribution 
was found to be small.
Experimental investigations show that introducing 6 dB overall 
losses -- in the best case equivalent to 20 km of
optical fiber -- leads to
an increment of only around 1\%,
hence to a QBER of 5-6\%. 
Besides detector noise, another major problem of all quantum key 
distribution schemes
developed up to now is stability, the only exception being \cite{plug&play}.
In order to really implement our setup for quantum cryptography,
the interferometers have to be actively stabilized, taking for instance
advantage of the free ports. The chosen passive stabilization
by controlling the temperature is not sufficient to ensure stable 
phases over a
long time.


To conclude, we presented a new setup for quantum cryptography using 
Bell states based on
energy-time entanglement, and demonstrated its feasability in a 
laboratory experiment. 
We found bit rates of around 33 Hz and
quantum bit error rates of around 4 \% which is low 
enough to ensure secure key distribution. Besides a smaller 
vulnerability 
against eavesdropper attacks, the advantage of using discrete 
energy-time states, up
to dimension 4, in our scheme, is the fact that no fast change 
between non-commuting bases 
is necessary. Nature itself chooses between the complementary 
properties 
energy and time. Furthermore, the recent demonstration that
energy-time entanglement can be preserved over long distances 
\cite{fullengthbell} 
shows that this scheme is perfectly adapted to long-distance 
quantum cryptography.

We would like to thank J.D. Gautier for technical support and PicoQuant 
for fast delivery of the laser. This work was supported by the Swiss PPOII
and the European QuCom IST projects.



\begin{table}
\begin{tabular}[5]{c|cccc}
&++&+--&--+&-- --\\
\hline
$s_Ps_A\&s_Ps_B$&278$\pm$6&197$\pm$5&187$\pm$5&147$\pm$4\\
$l_Pl_A\&l_Pl_B$&304$\pm$7&201$\pm$5&200$\pm$5&148$\pm$5\\
$s_Ps_A\&l_Pl_B$&11$\pm$0.8&10.4$\pm$0.8&9.2$\pm$0.7&9.4$\pm$0.7\\
$l_Pl_A\&s_Ps_B$&11.2$\pm$0.4&8.6$\pm$0.4&9.1$\pm$0.4&8.5$\pm$0.4\\
\hline
QBER [\%]&3.7$\pm$0.2&4.6$\pm$0.2&4.5$\pm$0.2&5.7$\pm0.3$
\label{time-basis}
\end{tabular} 
\caption{Results of the measurement in the time basis. 
The different coincidence count rates are 
due to different 
quantum efficiencies of the detectors, and the
slight asymmetry in the correlated events can be explained by
the non-equal transmission probabilities within the interferometers.}
\end{table}

\begin{table}
\begin{tabular}[5]{c|cccc}
&++&+--&--+&-- --\\
\hline
Visibility [\%]&92.5$\pm$1.8&92.6$\pm$1.4&89.3$\pm$1.9&94.5$\pm$1.6\\
max.&518$\pm$13&416$\pm$8&359$\pm$9&279$\pm$7\\
min.&20$\pm$5&16$\pm$3&20$\pm$4&8$\pm$2\\
\hline
QBER [\%]&3.7$\pm$0.9&3.7$\pm$0.7&5.3$\pm$1&2.8$\pm0.8$
\label{energy-basis}
\end{tabular}
\caption{Results of the measurement in the energy-basis.}
\end{table}


\begin{thebibliography}{99}
\bibitem{QNDattack} When using coherent states, the eavesdropper 
        can take advantage
        of the fact that there always is a small probability of finding 
        more than one photon in the same pulse, hence in the same state. 
        In these cases, she simply measures one and lets 
        pass the other one unmeasured. This
        strategy can be somewhat circumvented using photon pairs, where 
        one photon serves as trigger for the second one. 
        However, since the possibility to create 
        more than one pair at the same time increases with the 
        pump intensity, 
        we face a similar problem concerning the simultaneous presence 
        of more than one photon in the same state. Note that 
        this probability
        can be made arbitrarily small in both cases using weaker pulses or 
        lower pump energies. Still, there is an important difference,
        favouring the implementation of photon pairs: Whenever the trigger 
        photon has 
        been detected, there is one photon left in the setup. This 
        contrasts with
        the schemes based on faint pulses where the detecting electronics 
        has to be activated
        whenever a weak pulse (containing most of the time no photon) was 
        emitted.
\end{thebibliography}
\end{document}